# Network Coding Power Control Mechanisms for Time Varying Channels

Samah A. M. Ghanem
IEEE, Senior Member

*Abstract*—In this paper, we propose a model for large scale fading channels via markov process. We exploit the channel delay profile and the dependency between channel states via a first order autoregressive model that cast insight to the channel variations under fading and the closed form delay induced. We propose a network-coding structure that can be employed to compensate for the channel variations under fixed power and to the period of zero packet transmissions under adaptive power control. Satellite communications is an application to the model proposed.

## I. INTRODUCTION

Network Coding mechanisms have been adopted for the last few years since the milestone work of [1]. It has been few contributions since then that aims to model in a packet level the process of network coded transmission [2]. Further, the authors in [2] explored the timing nature of markov processes in the context of coding across packets over time division duplexing channels capitalizing on the time to absorption, see [3]. However, despite the novelty of the model proposed, the model considers channels with fixed erasure probabilities, which corresponds to channels at steady state taking apart the channel variation over time due to different fading processes. Therefore, it was of particular relevance to tackle the network-coding mechanisms over time varying channels and to describe on a packet level the statistical model of time-varying channels; the modeling process provides the physical prcoesses taking part in the propagation and reception process of a packet [4]. Therefore, statistical models are more customizable to real-world than empirical ones which are based on fitting of measured data, and so it can serve the purpose of this paper, aiming to describe, design, and optimize network-coding mechanisms on top of scenarios where fading causes time variation and attenuation over the transmitted signals. It is well known that the log-normal distribution is used to describe the variations of the modulated signal amplitude due to attenuation caused by obstacles impeding the line of site path. Therefore, we try in this paper to put forth a statistical model for large scale log-normal fading channels via a markov chain [1]. The variation of the channel states over time induces a time variation in each packet transmission, therefore, we exploit the channel delay profile and the dependency between channel states via a first order autoregressive model (AR1)[2] that casts insight to the channel variations under fading and consequently, we derive a closed form for the delay induced due to the fading process. The autoregressive model provides a general form to the markov model where each log-normally generated channel state(s) per markov state -with or without subcarrier correlation- puts forth the general distribution of the next states. We capitalize on the statistical properties of the markov process to derive a closed form expression for the time to absorption corresponding to the time to deliver N-coded packets under the fading process and in conjunction with a power control policy, therefore we provide optimal designs when coding across packets takes place.

The paper is organized as follows, section II introduces the channel model, Section III introduces the time-varying channel model in the packet level transmission, the expected time to deliver the packets, Section IV we introduce different power control policies with the corresponding outage probabilities. Section V introduces the network-coding scheme and the correlation framework, then we introduce optimal designs for delay and energy considerations. Finally, Section VI presents illustrative results.

## II. CHANNEL MODEL

Consider a downlink transmission over a Ka-band satellite channel, the received vector will be modeled by,

$$y(t) = h(t)P_T x(t) + n(t) \qquad (1)$$

$x$ and $y$ correspond to the transmit and receive symbols respectively. Fading due to rain in the Ka-band is well described as a slow time-varying, log-normally distributed process, with the channel gain between transmitter and receiver modeled by $h \sim log-N(m, \sigma^2)$, $P_T$ is the transmitted power, and $N$ is the zero mean complex white Gaussian noise. The mean of $h$ is $\mathcal{E}[h(t)] = e^{(m+\sigma^2)/2}$, and its variance is $\mathcal{E}[h(t)^2] = e^{(2m+\sigma^2)}(e^{\sigma^2} - 1)$. Therefore, the channel gain probability density function is given by,

$$p_h(h) = \frac{1}{h\sigma\sqrt{2\pi}} e^{-(20\log_{10} h - m)/2\sigma^2}, \qquad (2)$$

where $20\log_{10} h$ follows a Gaussian distribution. Hence, choosing $m = -\sigma^2/2$ makes the average power loss due to channel fading equals unity. And to ensure that the fading does not attenuate or amplify the average power, $h$ is normalized such that the moment generating function[3] $M_h(2) = \mathcal{E}[h(t)^2] = 1$. Therefore, a typical value is to choose $m = -0.5$, and $\sigma^2 = 1$ [7].

---

[1]The channel model is analyzed for the log-normal distributions, however, the same model can be analyzed for other small scale fading processing like Rayleigh, Ricean, or mixed ones.

[2]The autoregressive model can be of first or higher order, however AR1 is an appropriate choice for sattelite channels, see [5].

The dynamics of rain fading in the Ka band are generally described by an auto-regressive moving-average(ARMA) model with order $p$ [8], [9], [10] as follows,

$$h(t) = -\sum_{i=1}^{p} a_i h(t-i) + \omega(t) \quad (3)$$

$\omega$ is a white zero mean Gaussian process, and $a_i$ is the AR filter coefficients, see [10]

$$a_i = \frac{\mathcal{E}[h(t)h(t-i)]}{\sigma^2} \quad (4)$$

With $\mathcal{E}[h(t)h(t-i)]$ corresponds to the power delay profile (PDP) of the channel. Therefore, if $\sigma = 1$ then $a$ corresponds to the PDP of the channel. A special case of this model is the first-order autoregressive (AR1) model, which was found in [5] to provide a good match with experimental data. The AR1 model represents the fading process in discrete time as,

$$h(t) = -a_1 h(t-1) + \omega(t) \quad (5)$$

And the moving average appears in the above equation as follows,

$$h(t) = m - a_1 h(t-1) + \omega(t) \quad (6)$$

Thus, $a_1$ is a one-step autocorrelation coefficient,

$$a_1 = \frac{\mathcal{E}[h(t)h(t-1)]}{\sigma^2} = \frac{R_h(T_s)}{R_h(0)} \quad (7)$$

Where $R_h(t)$ corresponds to the autocorrelation function of the underlying continuous-time process. The sampling period is $T_s$. The corresponding power spectral density (PSD) of the AR1 model has the following rational form,

$$S_h(f) = \frac{\sigma^2}{|1 + a_1 e^{-j2\pi f}|^2} = \frac{\sigma^2/\pi f_d}{1 + (f/f_d)^2} \quad (8)$$

With $a_1 = e^{-\beta}$, with $\beta = 2\pi f_d T_s$ being the normalized fading rate through which the power delay profile decays exponentially with $f_d$ the Doppler spread of the fading process.

After generating the channel vector $h$ corresponding to the log-normally distributed channel gains at a certain markov state, the estimation of the channel gains at one-step forward transition can be found by solving the Yule Walker equations to find the AR filter coefficients, multiplying equation (5) by $h(t-1)$ and taking the expectation of both sides we have,

$$\mathcal{E}[h(t)h(t-1)] = \mathcal{E}[h(t-1)m] - \mathcal{E}[a_1 h(t-1)^2] + \mathcal{E}[h(t-1)\omega(t)] \quad (9)$$

Therefore, the equation above becomes as follows,

$$R_h(t) = \mathcal{E}[h(t)h(t-1)] = \mathcal{E}[a_1 h(t-1)^2] = a_1 \sigma^2 \quad (10)$$

[3]The moment generating function of $h$ is given by $M_h(s) = exp(m_h s + \frac{\sigma_h^2 s^2}{2})$, [6].

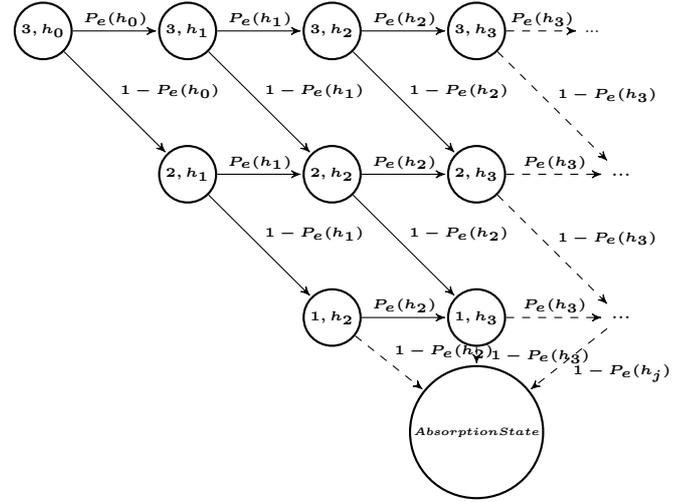

Figure 1. Channel State Autoregressive Model with First Order Dependency: 3-Packets Transmission

## III. TIME-VARYING CHANNEL MODEL OF PACKET TRANSMISSION

Figure 1 illustates a packet level modeling for time-varying channels, each packet transmission corresponds to a vector of transmitted symbols that suffer from fading over time, each vector is represented by one state in the Markov chain. Therefore, we first consider one channel coefficient corresponding to one packet transmission, however, digging into the physical layer and the bit- or symbol- level modeling, the sum of log-normally distributed channel states can be represented by one log-normally distributed random variable which corresponds to the packet level. To transmit N-packets, each packet transmission will be either successful, keeping N-1 packets to be transmitted at a different channel state, or will fail, making the process to be repeated at a different channel state until all N-packets are successfully received.

### A. The expected time to transmit one packet under small scale fading

We will first analyse the model with no network-coding on top of it. This will make it easier to understand how to generate redundancy at the packet level and how to combine it with feedback later on. Consider that the channel states are dependent to all possible permutations of the past states. Therefore, we can write the expected time to deliver i-packets in the following closed form,

$$T(i, h_j) = T_d(i, h_j) + \sum_{\forall l,k} p_{(i,h_j) \to (l,h_k)} T(l, h_k) \quad (11)$$

Figure 1 illustrates the autoregressive markov model with order $p = 1$ through which the current channel state is only dependent on one-time lag past state -Note that this is a fair assumption when considering practical systems like satellite wideband systems- [8]. Therefore, the expected time to deliver i-packets becomes as follows,

$$T(i, h_j) = T_d(i, h_j) + p_{(i,h_j) \to (i-1,h_{j+1})} T(i-1, h_{j+1})$$
$$+ p_{(i,h_j) \to (i,h_{j+1})} T(i, h_{j+1}) \quad (12)$$

The time to deliver a packet at a given channel state equals the packet length $T_d(i, h_j) = T_p$, and the transition probability $p_s$ corresponding to successful transition of one packet $p_{(i,h_j) \to (i-1,h_{j+1})}$ is given by,

$$p_{(i,h_j) \to (i-1,h_{j+1})} = (1 - p_e(h_j)) \quad (13)$$

And the probability of failure $1 - p_s$ in transmitting one coded or uncoded packet $p_{(i,h_j) \to (i,h_{j+1})}$ is given by,

$$p_{(i,h_j) \to (i,h_{j+1})} = p_e(h_j) \quad (14)$$

Where $p_e(h_j)$ is the packet erasure probability at a given state which is channel dependent. Therefore, we can write the time evolution of the packet transmission over the channel fading states in a matrix form $T$, where the matrix $T$ is not necessarily symmetric, as follows,

$$T = \begin{pmatrix} T_{1,h_1} & T_{1,h_2} & \cdots & T_{1,h_n} \\ T_{2,h_1} & T_{2,h_2} & \cdots & T_{2,h_n} \\ \vdots & \vdots & \ddots & \vdots \\ T_{N,h_1} & T_{N,h_2} & \cdots & T_{N,h_n} \end{pmatrix}$$

It is straightforward to think that the expected time to deliver N-packets can be the discrete sum of the packet transmission over different channel states, however, given a packet transmission and retransmission, an N-packet transmission will be a function of the previous N-1 packet transmissions and so the channel states differ in the transmission-retransmission parts. Emphasizing that the model considers that a packet cannot be retransmitted at the channel state that it was transmitted with, therefore we account for the variations in the fading process, and so the time evolution clearly depicted numerically, i.e., $T(N, h_j)$ is greater than $T(N, h_{j-1})$, and $T(N, h_j)$ is greater than $T(N-1, h_j)$, and so on and so forth.

## IV. Network Coding Scheme

We will propose a network-coding structure that emphasizes the model proposed which can account for channel variability over time. In particular, for one state transmitting $N_i$ coded packets that would account for the channel variations and power adaptation as well as the degrees of freedom of the receiver. If the $N_i$ coded packets are transmitted under certain channel variability, received and decoded successfully, it will be sucessfully acknowledged, if $i - j$ coded packets received and acknowelged, the transmitter will need to transmit or retransmit the $N_i - i + j$ coded packets failed to be received at the first transmission, so they will be transmitted with new channel states, the process is repeated until all coded packets that are adaptively not transmitted due to channel variability are compensated with encoding across the packets. Therefore, lets give a new name to the state and call it a super state where $N_i$ coded packets need to be transmitted underlined by all the states in the markov chain above. It follows that there are two transition probability components which are, $p_{(i,h_i) \to (i,h_k)}$ and $p_{(i,h_i) \to (k,h_k)}$ correspond to the main transitions between super states, and given by,

$$p_{(i,h_j) \to (i,h_k)} = p_{(i,h_j) \to (i,h_{j+1})} p_{(i,h_{j+1}) \to (i,h_{j+2})} \cdots$$
$$\cdots p_{(i,h_{k-1}) \to (i,h_k)} \quad (15)$$

Therefore, the expected time to deliver $N_i$ coded packets will be a function of the expected time to deliver the packets at their given -or predicted- is as follows,

$$T(i, h_j) = T_d(N_i, h_j) + \sum_{l=1}^{i} P^{N_i}_{(i,h_j) \to (l,h_{j+N_i})} T(l, h_{j+N_i+1}), \quad (16)$$

with $T_d(N_i, h_j) = (N_i + 1) T_p$ where the acknowledgment time appears as an addition of one into the time slot indices in the equation above, and the round trip time is considered to be zero. The matrix $P$ is the one step transition matrix of the model proposed, however,

$$\left( \prod_{i=1}^{N_i} P \right)_{(i,h_j) \to (l,h_{j+N_i})} = P^{N_i}_{(i,h_j) \to (l,h_{j+N_i})}, \quad (17)$$

corresponds to all transition probabilities over the time slots from the initial given log-normally distributed channel, with state $h_j$ until the estimated state at $j + N_i$ time slot. The $j + N_i + 1$ appears in the timing consideration due to the acknowledgment.

### A. Delay-Energy Optimization Tradeoff

Minimizing the delay requires different design aspects optimizing over the optimal number of coded packets to be transmitted with respect to the receiver degrees of freedom, however, to minimize the energy per transmission, either optimal power control need to be considered, or otherwise an acceptable tradeoff between delay and energy considerations need to take place.

*1) Minimum time to deliver N-coded packets:* The minimum expected time to deliver N-coded packets can be optimized to find the optimal number of coded packets to transmit over a single transmission, as follows,

$$min_{N_1,...,N_i} T(i, h_j) = min_{N_1,...,N_i} T_d(N_i, h_j) +$$
$$min_{N_1,...,N_i} \sum_{l=1}^{i} P^{N_i}_{(i,h_j) \to (l,h_{j+N_i})} T(l, h_{j+N_i+1}), \quad (18)$$

While its not possible to find a closed form expression for the optimal number of coded packets, one can find the optimal $N_1^*$ [2] for a fixed erasure probability. However, such a problem was addressed and solved via adaptive network coding [11], [12], [13].

*2) Minimum energy used to deliver N-coded packets:* The energy required to deliver N-coded packets is the sum over the time required to deliver those packets multiplied by the transmission power $P_t$ used in each transmission, therefore, we aim to minimize the energy as follows,

$$min_{P_{t_1},...,P_{t_{N_i}}} E(N_i, h_j) =$$
$$min_{P_{t_1},...,P_{t_{N_i}}} E(i, h_j)+$$
$$min_{P_{t_1},...,P_{t_{N_i}}} \sum_{l=1}^{i} P^{N_i}_{(i,h_j) \to (l,h_{j+N_i})} E(N_l, h_{j+N_i+1}), \quad (19)$$

Which can be written as,

$$min_{P_{t_1},...,P_{t_{N_i}}} P_{t_{N_i}} T(N_i, h_j) =$$
$$min_{P_{t_1},...,P_{t_{N_i}}} P_{t_{N_i}} T(i, h_j)+$$
$$min_{P_{t_1},...,P_{t_{N_i}}} \sum_{l=1}^{i} P^{N_i}_{(i,h_j) \to (l,h_{j+N_i})} P_{t_{N_l}} T(N_l, h_{j+N_i+1})$$
(20)

If the power is fixed over all transmissions $P_{t_1} = P_{t_1} = ... = P_{t_{N_i}}$, then we can only optimize over the number of coded packets, and if an adaptive power control is used we can employ both optimal strategies together. Moreover, minimizing the probability of error over transmitted power is another way to tackle both problems to have an adaptive optimal power control. Therefore, in the following section we will analyze a set of power control policies and their implication on the delay, and when coding across packets is employed.

## V. POWER CONTROL POLICY AND OUTAGE PROBABILITY

We will analyze the model under the assumption of uncorrelated dependent channels, when a fixed power strategy over all transmissions is implemented, and if an adaptive power control strategy based on the channel variation over each packet transmission is implemented, and we will introduce the implication of the strategies on the bit error outage probability. Therefore, we will first consider that the channels that underline each state of the markov chain is uncorrelated, such that the probability density function for the log-normally distributed channel channel $p(h)$ defined as in (2).

### A. Fixed Power

We first examine the conventional, fixed margin link design with a fixed power $P_t$ used over all transmissions, it follows that the received singal to noise ratio varies accordignly with the channel gain.

Consider the upper bound on the bit error probablity given by,

$$P_b(h_j) \leq Q(\frac{m - ln(h_j)}{\sigma}) = \frac{1}{2} erfc(\frac{m - ln(h_j)}{\sqrt{2}\sigma}) \quad (21)$$

The signal to noise ratio $\gamma = \frac{P_t h}{N_0 R} = SNRh$ where $N_0$ is the noise power spectral density and $R = \frac{1}{T_b}$ is the baud rate given by the inverse of the time of the information bit. Therefore, the bit error rate probability can be written as a function of the SNR as follows,

$$P_b(h) \leq Q(\frac{m - ln(\frac{h}{SNR})}{\sigma}) = \frac{1}{2} erfc(\frac{m - ln(\frac{h}{SNR})}{\sqrt{2}\sigma}) \quad (22)$$

Moreover, the packet erasure probability at channel state $h_j$ with respect to the bit error probability is given by,

$$p_e(h_j) = 1 - (1 - P_b(h_j))^B \quad (23)$$

$B$ is the total number of bits in a packet.

### B. Adaptive Power Control

In the adaptive power control strategy, the transmitted power $P_t$ can be varied accordingly with the known fading gain so that the SNR is maintained at a certain value. Therefore, as mentioned in [8], an appropriate power adaptation policy is as follows,

$$\begin{cases} P_t \frac{h_{out}}{h}, & P_t \frac{h_{out}}{h} \leq P_{t_{max}} \\ 0 & , otherwise \end{cases} \quad (24)$$

The outage channel state $h_{out}$ is given with respect to the outage probability given in [8] as,

$$P_{out} = Q(\frac{m - ln(h_{out})}{\sigma}) = \frac{1}{2} erfc(\frac{m - ln(h_{out})}{\sqrt{2}\sigma}) \quad (25)$$

Moreover, the bit error probability given the power adaptation policy is used will be as follows,

$$P_b(h) \leq Q(\frac{m - ln(\frac{h_{out} N_0 R}{P_t})}{\sigma}) = \frac{1}{2} erfc(\frac{m - ln(\frac{h_{out} N_0 R}{P_t})}{\sqrt{2}\sigma}) \quad (26)$$

Therefore, for the adaptive power policy, maintaining a specific SNR means that the time model maintains the same probability of bit error rate and so the same probability of erasure (while there is no outage), however, when an outage is detected, the erasure probability becomes 1 (meaning that our policy should not transmit. This causes a zero transmission in certain times. Therefore, a mixed policy of adaptive power control or fixed power could lie underneath a network-coded framework where the usage of coding across the transmitted packets can compensate for the times when the adaptive strategy is implemented, therefore, a network-coded scheme that could adapt to the software-hardware of the system; wether it is a fixed power transmission or an adaptive power.

### C. Power Control and Outage with Network Coding

Coding across packets introduces one kind of correlation across the packets, therefore, in the following analysis we consider that the channels that underlie each state of the markov chain are correlated, such that the probability density function is the probability distribution of the sum of log-normally distributed correlated subchannels[4] with $h_j = \sum \omega_1 h_{1,j} + ... + \omega_{N_i} h_{N_i,j} \quad \forall \quad h_j$ corresponds to each state in the markov chain. And the moment generating function is given by Eq.6, [14].

---

[4]The assumption of a sum of log-normally distributed random variables of the channel coefficients is a simplification of the reality where a weighted sum is the usual case when random network coding across packets is used, or from the bit level perspective, each weight will be decided by the modulation and the transmission power used.

The probability density of the sum of log-normally distributed correlated random variables corresponding to subcarrier correlation can be given by,

$$p(h_j = \sum_{j=1}^{N_i} h_{i,j}) = \frac{1}{(2\pi)^{N_i/2}|C|^{1/2}} \prod_{j=1}^{N_i} e^{\frac{-(ln(h_j)-\mu)^\dagger C^{-1}(ln(h_j)-\mu)}{2}}$$ (27)

The correlation matrix $C$ can be given by,

$$C = \begin{pmatrix} 1 & \rho_{1,2} & \cdots & \rho_{1,k} \\ \rho_{2,1} & 1 & \cdots & \rho_{2,k} \\ \vdots & \vdots & \ddots & \vdots \\ \rho_{k,1} & \rho_{k,2} & \cdots & 1 \end{pmatrix}$$

Consequenctly, we can analytically derive the upper bound of the bit error probability for higher dimensions, where all possible correlation permutations over the $k-$dimensions of correlations can be estimated and we integrate over all correlated channel states,

$$P_b(h_{1,j}, h_{2,j}, ...., h_{N_i,j}) \leq Q(\frac{m - ln(h_{1,j})}{\sigma}, \frac{m - ln(h_{2,j})}{\sigma}, ..., \frac{m - ln(h_{i,j})}{\sigma}, C)$$ (28)

With the $n-$dimensional Gaussian Q-function defind by,

$$Q(X = [x_1, x_2, ..., x_n], C) =$$
$$\frac{1}{(2\pi)^{N_i/2}|C|^{1/2}} \int_{x_1}^{\infty} \int_{x_2}^{\infty} .... \int_{x_n}^{\infty} \prod_{j=1}^{N_i} e^{\frac{-(X-\mu)^\dagger C^{-1}(X-\mu)}{2}} dx_1 dx_2...dx_n$$ (29)

For the simple case of 2x2 correlation matrix we have,

$$C = \begin{pmatrix} 1 & \rho_{1,2} \\ \rho_{2,1} & 1 \end{pmatrix}$$

With $\rho_{1,2} = \rho_{2,1} = \rho$, and $\rho_{i,i} = 1$. We can derive the bit error rate outage probability as follows[5],

$$P_b(h_{1,j}, h_{2,j}) \leq Q(\frac{m - ln(h_{1,j})}{\sigma}, \frac{m - ln(h_{2,j})}{\sigma}, \rho)$$ (30)

With the two-dimensional Gaussian Q-function defind by,

$$Q(x_1, x_2, \rho) = \frac{1}{2\pi\sqrt{1-\rho^2}} \int_{x_1}^{\infty} \int_{x_2}^{\infty} e^{(-\frac{u^2+v^2-2\rho uv}{2(1-\rho^2)})} du dv$$ (31)

Notice that the correlation matrix can be decomposed into a unitary matrices $U$ and $\Sigma_c$ the covariance matrix of the subcarrier gains, which decomposes into a Fourier matrices $F$, and the covariance matrix $\Sigma_h$ of the channel vector $h$ at the markov state $i$ which corresponds to the power delay profile induced due to the variation of the channel states.

$$C = U\Sigma_c U^\dagger = UF\Sigma_h F^\dagger U^\dagger$$ (32)

Therefore, the network coding can be analyzed under different policies when fixed power is used and when adaptive power is used with correlation across the transmitted coded packets or when a transmit diversity and de-correlation techniques are used like precoding which can be employed to diagonalize the channel if multiple antennas are available and also to decorrelate the inputs. In particular, we can employ the following precoding structure,

$$P = U_p \Sigma_p V_p^\dagger = UF diag(P_t) V_p^\dagger$$ (33)

[5]Notice that this result matches with the one found in [15] for selection and stay combining (SSC) technique used at the reciever side, for instance, for the last and most simple closed form obtained with $\mu_1 = \mu_2 = \mu$ and $\sigma_1 = \sigma_2 = \sigma$, and correlation coefficient $\rho$.

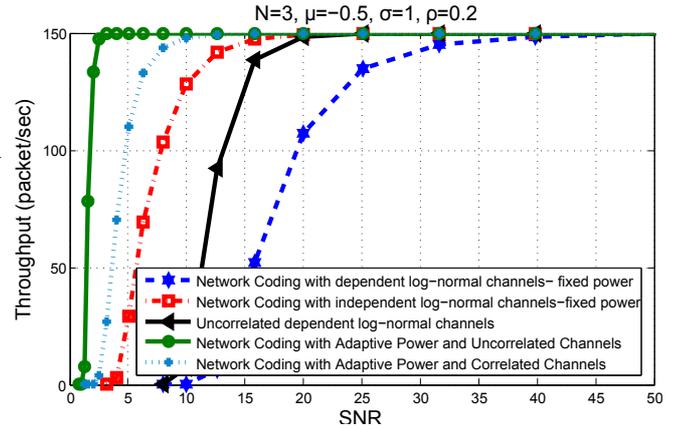

Figure 2. Throughput vs. SNR

With $V_p$ can be chosen in a way such that it works as a rotation matrix to insure firstly, allocation of power into the strongest channel singular vectors, and secondly, to diagonalize the channel insuring uncorrelated error or in other words independency between inputs, see [16]. Henceforth, the autoregressive feature of the preceding states will be derived the same way as we expained at the begining where correlation can be considered between different dependent states, then the upper bound on the bit error proability and so the erasure probability can be evaluated to derive the expected time to deliver the N-packets.

## VI. SIMULATION RESULTS

We shall now present a set of results to provide further insight into the problem. We consider a packet length $T_p = 1/150$ sec with fixed power policy is used $P_t = 1$ and the log-normally generated channels are with mean $m = -0.5$ and variance $\sigma = 1$, we first analyze the delay-throughput-erasure probabilities with respect to the SNR for the case where no autoregressive model exists, and therefore the channels are log-normally distributed independent channels, secondly we will compare this to the case when first order dependency exists between the channels, and thirdly, when network-coding is done jointly with the power control policies, therefore, correlation is added by having weighted sums by coding across the packets, and we assume one to one correlation, i.e. two dimensional sum of correlated lognorms $\rho = 0.2$. Notice that for the adaptive power policy without network coding, its straightforward to know that a fixed delay-throughput can be reached maintaining the SNR, for a given outage probability.

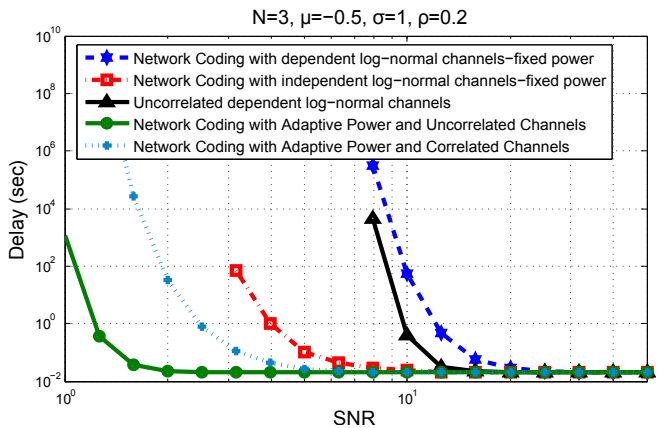

Figure 3. Time to deliver N-packets vs. SNR

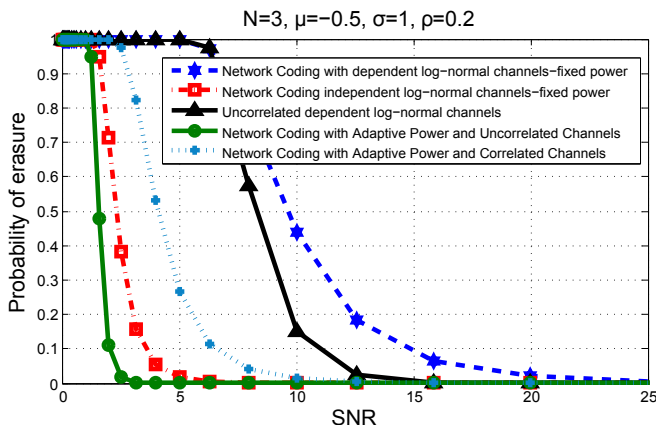

Figure 4. Probability of erasure vs. SNR

Figure2 clearly depicts the maximum throughput reached in each case which is equal to the $1/T_p$, due to the channel delay profile enduced via variation, dependency or dependency with correlation, this rate is approached at different high-SNR margins, while at the low-SNR the behavior grows based on the three different aspects.

Intuitively, we can understand the increase in delay in Figure3 -therefore higher erasures, and lower throughput- for the dependent channels in comparison to the independent ones. However, on the one hand, its worth to notice that employing network coding in a fixed power strategy causes extra delay or in other words we paid a time cost due to the induced correlation. On the other hand, network-coding gives a better solution with the adaptive power strategy, where not only lost but also silent periods of no transmissions will be compensated, however, it provides another kind of adaptivity where the transmitter can adapt to the receiver experience, and using the precoding structure that can diagonalize the channels and decorrelate the inputs can be a solution that can move the network-coding to be optimal for the adaptive policy. Figure 4 shows a straightforward result emphasizing the previous ones where maintaining a least of 10 dB SNR significantly improves the throughput and delay performance. Intuitively, we can understand the increase in delay in Figure3 -therefore higher erasures, and lower throughput- for the dependent channels in comparison to the independent ones. However, on the one hand, its worth to notice that employing network coding in a fixed power strategy causes extra delay or in other words we paid a time cost due to the induced correlation. On the other hand, network-coding gives a better solution with the adaptive power strategy, where not only lost but also silent periods of no transmissions will be compensated, however, it provides another kind of adaptivity where the transmitter can adapt to the receiver experience, and using the precoding structure that can diagonalize the channels and decorrelate the inputs can be a solution that can move the network-coding to be optimal for the adaptive policy. Figure 4 shows a straightforward result emphasizing the previous ones where maintaining at least 10 dB SNR can significantly improve the throughput and delay performance.

## VII. CONCLUSION

We have addressed the problem of delay in time varying channels, we analyzed two power allocation policies from an outage probability perspective, and based on the costs paid via a fixed power allocation policy associated with higher packet losses, and the power costs paid using an adaptive power policy is another tradeoff. However, due to channel variation, there is a time where the adaptive policy dictates no transmission to occur when outage probability is 1. Therefore, we proposed a network-coded scheme that can mitigate the channel variations over time compensating for the power cost as well as silent transmission in few times, therefore, we may pay some extra delay on the encoding/decoding process causing correlation between inputs over dependent channels. However, we propose some optimal designs that can work in conjunction with the network coding; the optimal decision adapts to the hardware avaialble at the transmitter where decorrelating and transmit diversity techniques can be employed with the power allocation, while it also adapts to the receiver experience where the transmission and coding across packets can adaptively chose the optimal number of coded packets to transmit based on the degrees of freedom of the receiver as well as the receiver diversity techniques used.